\documentstyle [12pt,epsf]{article}

\newcommand{\beq}{\begin{equation}}
\newcommand{\dd}{\partial}
\newcommand{\eeq}{\end{equation}}
\newcommand{\bea}{\begin{eqnarray}}
\newcommand{\eea}{\end{eqnarray}}
\newcommand{\F}{\Phi}
\newcommand{\f}{\phi}
\newcommand{\vf}{\varphi}
\newcommand{\Q}{\tilde{Q}_{_L}}
\newcommand{\q}{\tilde{q}_{_R}}
\newcommand{\Lp}{\tilde{L}_{_L}}
\newcommand{\lp}{\tilde{l}_{_R}}
\newcommand{\e}{{\cal E}_\omega}

\begin{document}
\baselineskip 7.4 mm

\def\thefootnote{\fnsymbol{footnote}}

\begin{flushright}
\begin{tabular}{l}
CERN-TH/97-70 \\
hep-ph/9704273 \\
April, 1997 
\end{tabular}
\end{flushright}

\vspace{2mm}

\begin{center}

{\Large \bf 
Solitons in the supersymmetric extensions of the Standard Model 
}
\\ 
\vspace{8mm}

\setcounter{footnote}{0}

Alexander Kusenko\footnote{ email address:
kusenko@mail.cern.ch} \\
Theory Division, CERN, CH-1211 Geneva 23, Switzerland \\

\vspace{12mm}

{\bf Abstract}
\end{center}

All supersymmetric generalizations of the Standard Model
allow for stable non-topological solitons of the Q-ball type which may 
have non-zero baryon and  lepton numbers, as well as the electric charge.  
These solitons can be produced in the early Universe, can affect the 
nucleosynthesis, and can lead to a variety of other cosmological 
consequences.

\vfill

\pagestyle{empty}

\pagebreak

\pagestyle{plain}
\pagenumbering{arabic}
\renewcommand{\thefootnote}{\arabic{footnote}}
\setcounter{footnote}{0}

\pagestyle{plain}

Supersymmetric generalizations of the Standard Model (SSM)  involve 
a complicated scalar potential that depends on a large
number of variables.  Although the details of such a potential 
depend on a model, a generic feature of all SSM is the
presence of the tri-linear couplings  
of the type $H \F \f$, where $\F$ is a left-handed squark ($\Q$) or slepton
($\Lp$) doublet, and $\f$ is the corresponding right-handed singlet 
($\q$ or $\lp$) of the SU(2).  These terms arise from the Yukawa
couplings in the superpotential, as well as from the supersymmetry
breaking terms.  We will show that such cubic interactions lead to the
appearance of non-topological solitons in the spectrum of the SSM.  
Solitons of this type, dubbed {\it Q-balls} \cite{coleman1}, can have 
a non-zero baryon or lepton number, or electric charge. 
They can lead to interesting cosmological consequences and may provide new
constraints on the parameters of the SSM.   

We argue that $B$ and $L$ balls created in the early Universe can also play
an important role in the synthesis of nuclei by producing lumps of nuclear
matter prior to the onset of the standard nucleosynthesis.  This opens a
new possibility for the production of heavy elements through fission of the
quark matter lumps that are left over after the decay of the 
squark and slepton Q-balls.  

It was shown \cite{ak_another} that 
very small Q-balls (Q-beads) with charges $Q \sim 1$ can 
exist, despite the fact that the usual thin-wall
approximation breaks down for small $Q$.  
A new formalism \cite{ak_another} that has been developed to analyse these
solitons gives an adequate description of Q-beads as long as the charge
and the tri-linear couplings in the potential are sufficiently small. 
Such small-charge
solitons are of particular interest for the phenomenology of the
MSSM, because the leptonic and baryonic beads can be absolutely stable due
to a combination of several conservation laws. 
They could be produced in large quantities in the early Universe and can
contribute to dark matter. 

Finally, a $B\neq 0$, $L\neq 0$ soliton interacts as a leptoquark,   
which has intriguing implications.

\section{Q-balls with many flavors}

We begin with a straightforward generalization of Coleman's discussion
of Q-balls \cite{coleman1} to the case that involves several scalar fields
with different charges.
 
Let us consider a field theory with a scalar potential $U(\vf) \equiv 
U(\vf_1,...,\vf_n)$ which has a global minimum
at $\vf=0$; $U(0)=0$.  
Let $U(\vf)$ have an unbroken global U(1) symmetry at the origin, 
where $\vf=0$.  The scalar fields $\vf_i$ have charges $q_i $
with respect to this $U(1)$, and at least one of $q_i \ (i=1,...,n)$ is not
equal to zero. 

The charge (taken to be positive for definiteness) of some field
configuration $\vf(x,t)$ is  

\beq
Q=\sum_k \, q_k \frac{1}{2i} 
\int \vf_k^* \stackrel{\leftrightarrow}{\partial}_t 
\vf_k \, d^3x
\label{Qt}
\eeq

Clearly, a configuration $\vf(x,t)\equiv 0$ has zero charge, so the
solution that minimizes the energy

\beq
E=\int d^3x \ \left [ \frac{1}{2} \sum_k |\dot{\vf_k}|^2+
\frac{1}{2} \sum_k |\nabla \vf_k|^2 
+U(\vf) \right], 
\label{e}
\eeq
and has a given charge $Q>0$, must differ from zero in some (finite) domain. 
We will use the method of Lagrange multipliers to look for the minimum of 
$E$ at fixed $Q$.  One must find an extremum of 

\begin{eqnarray}
\e & =& E+  \omega \left [ Q- \sum_k \, q_k \frac{1}{2i} 
\int \vf_k^* \stackrel{\leftrightarrow}{\partial}_t 
\vf_k \, d^3x
\right ]     
\label{Ew1} \\
& & \nonumber \\
& = & \int d^3x \, \frac{1}{2} \sum_k \left | \dd_t \vf_k 
- i \omega q_k \vf_k 
\right |^2 \ + \ \int d^3x \, \left [\frac{1}{2} \sum_k |\nabla \vf_k |^2  
+ \hat{U}_\omega(\vf)
\right ] + \omega Q ,
\label{Ewt}
\end{eqnarray}
where $ \omega$ is a Lagrange multiplier, and   

\beq
\hat{U}_\omega (\vf) = U(\vf)\ - \ \frac{1}{2} \omega^2 \, 
\sum_k q_k^2 \, |\vf_k|^2. 
\label{Uhat}
\eeq
Variations of $\vf(x,t)$ and
those of $\omega$ can now be treated independently, the usual advantage of 
the Lagrange method.

We are looking for a solution that extremizes $\e$, while all the physical
quantities, including the energy, $E$,  are time-independent.  
To minimize the first term in equation (\ref{Ewt}), the only one that  
appears to depend on time explicitly, one must choose

\beq
\vf_k(x,t) = e^{iq_k \omega t} \vf_k(x),
\label{tsol}
\eeq
where $\vf_k(x)$ is real and independent of time.  
We conclude that Q-balls with many flavors are solitons built of fields
that rotate in the internal space with velocities proportional to their
charges.  For the solution (\ref{tsol}), equation (\ref{Qt}) yields 

\beq
Q= \omega \sum_k q_k \int \vf_k^2(x) \ d^3x
\label{Qw}
\eeq

It remains to find an extremum of the functional 

\beq
\e = \int d^3x \, \left [\frac{1}{2} \sum_k |\nabla \vf_k(x) |^2  
+ \hat{U}_\omega(\vf(x))
\right ] + \omega Q ,
\label{Ew}
\eeq
with respect to $\omega$ and the variations of $\vf(x)$ independently.
We can first minimize $\e$ for a fixed $\omega$, while varying the shape of
$\vf(x)$.  The solution to this part of the problem \cite{ak_another} 
is just a bounce $\bar{\vf}_\omega (x)$ associated with tunneling in 
$d=3$ Euclidean dimensions \cite{tunn0,tunn,cgm,linde} in the potential
$\hat{U}_\omega (\vf)$.  The problem is, therefore, reduced to that which
is more familiar and better developed. This analogy was used in
Ref. \cite{ak_another}  to prove the existence and the classical stability
of the solitons in the limit of small charge.  For large $Q$, the
existence proof was given in  Ref.~$\cite{coleman1}$.  
From Ref.~\cite{cgm} we know that the solution is spherically symmetric:
$\bar{\vf}(x)=\bar{\vf}(r), \ r=\sqrt{\vec{x}^2}$. This implies, in
particular, that the ground state soliton has zero angular momentum. 

For a Q-ball to exist,  the following condition ({\it cf.}
Ref. \cite{coleman1}) must be satisfied:  

\beq
\mu^2=
2 U(\vf) \left/ \left (\sum_k q_k \vf_{k,0}^2 \right ) \right. = {\rm min},
\ \ {\rm for} \ |\vec{\vf}_0|^2 > 0.
\label{condmin}
\eeq
As discussed below, if $\ U(\vf) \left/ \left (\sum_k q_k \vf_{k,0}^2 
\right )\right.$ has a global minimum at $\vf_k=\vf_{k,0} \neq 0$, then 
Q-balls are stable with respect to decay into the $\vf$ quanta.  However,
if condition (\ref{condmin}) is satisfied in the sense of a local minimum,
then the corresponding soliton is metastable and can either dissociate into 
$\vf$ particles through tunneling, or evolve into a different soliton with
lower value of $\mu$.

\section{Thin-wall approximation for large Q-balls}

For clarity, in this section we assume that $\mu^2$ has only one minimum. 
Relaxing this constraint 
is straightforward and amounts to allowing Q-balls of different radii made
of different subsets of fields to overlap.  In some sense, this 
defines an ``irreducible'' Q-ball and will simplify the
algebra. 

For large $Q$, the solution that minimizes the energy can be approximated
\cite{coleman1} by a thin-wall ball of charged matter with a radius $R$:
$\vf_i(r) \approx \vf_0 \theta(R-r)$.  (Note that we use a single radius 
$R$ for all flavors, which is the simplification  due to restricting
our discussion to irreducible Q-balls only.)  One can eliminate
$\omega$ from the expression for the energy using  constraint (\ref{Qw})
and minimize $E$ with respect to the  volume $ V = 4 \pi R^3/3$ of
the soliton.  

\beq
E\approx \frac{Q^2}{2 (\sum_k q_k \vf_{k,0}^2) V}
+ U(\vf_0) V + {\rm surface \ energy \ (neglected)} \ = \ min  
\eeq
for $V \equiv 4 \pi R_0^3/3 = Q/\sqrt{2 U(\vf_{k,0})(\sum 
q_k \vf_{k,0}^2)}$ and  

\beq
M_{_Q}= E_{min}=Q \sqrt{2U(\vf_0) \left / \left (\sum_k q_k \vf_{k,0}^2
\right) \right.} 
\label{ElargeQ}
\eeq

The energy per unit charge,  $M_{_Q}/Q \approx \sqrt{2U(\vf_0)/(\sum_k q_k 
\vf_{k,0}^2)}$, is less than the mass of the lightest of the 
$\vf_k$ particles,  if condition (\ref{condmin}) is satisfied in the 
{\it strong} sense: that is if the minimum is global.  In this case, the 
Q-ball is  stable with respect to its decay to $\vf$ particles.

For large $Q$, the surface energy is small and can be neglected.  For
smaller $Q$, the surface energy becomes more important.  A naive
application of the thin-wall formalism seems to imply that only the $Q$
balls with a large enough charge, $Q>Q_{min}$, can exist.  This constraint,
however, is merely an artifact of the thin-wall approximation.
The latter fails
to account correctly for the energies of the wall and the interior when
they become inseparable, that is in the ``thick-wall'' case.  Q-balls
of small charges have been proven to exist \cite{ak_another}.  
There is no classical lower limit of the charge $Q$ of a (classically) 
stable Q-ball.  However, quantum consistency requires charge quantization
in units of the charge of the $\vf$ field.  Therefore, $Q\ge 1$.  Also, 
in the limit $Q\rightarrow 1$, quantum corrections can significantly modify
semiclassical results (at least, we do not have a proof to the contrary
\cite{ak_another}).

\section{Beyond the thin-wall approximation: Q-beads}

If $Q$ is small, $\omega$ becomes large\footnote{
This is not in contradiction with equation (\ref{Qw}). As $\omega$
increases, $(\int \bar{\vf}^2_\omega)$ for the bounce $\bar{\vf}^2_\omega
(x)$ in the potential $\hat{U}_\omega (\vf)$ decreases faster than 
$(1/\omega)$; see discussion in Ref. \cite{ak_another}.} 
\cite{ak_another}.  For large $\omega$, the bounce in the potential 
$\hat{U}_\omega(\vf)$ cannot be analysed using the thin-wall approximation. 
A ``thick-wall'' approximation \cite{linde,kls} can be used instead. 
We will briefly summarize the results of Ref.~\cite{ak_another} relevant to
our discussion. 

For a single scalar field with a potential\footnote{The $\vf^3$ term
should be thought of as a $U(1)$-symmetric cubic interaction, {\it e.\,g.},
$(\vf^\dag  \vf)^{(3/2)}$.  In the MSSM, the tri-linear couplings of the
Higgs field to squarks and sleptons yield the requisite cubic terms, 
whose ``flavor structure'' is discussed in the next section.} $U(\vf) =
\frac{1}{2} M^2 \vf^2 - A \vf^3 + \lambda_4 \vf^4$, one has to calculate
the bounce in the effective potential 

\beq
\hat{U}_\omega (\vf)= \frac{1}{2}(M^2-\omega^2) \vf^2 - A \vf^3 + 
\lambda_4 \vf^4 
\label{U234}
\eeq
and then minimize $\e$ in equation (\ref{Ew}) with respect to $\omega$. 
The thick-wall approximation \cite{ak_another} is applicable and the  
minimum exists if 

\beq
Q  \ll  \frac{3 S_\psi M} {A} \times {\rm min} \left
( \frac{1}{\sqrt{\lambda_4}}, \frac{M}{2A} \right )
\label{Qconstr}
\eeq  
where $S_\psi \approx 4.85$.  The small $Q$ soliton has a mass $M_{_Q}$ and
a size $R_{_Q}$:

\begin{eqnarray}
M_{_Q} & \approx &  
Q M \ \left [  1-\frac{1}{6} \epsilon^2 - \frac{1}{8} \epsilon^4 \ -
 \  O(\epsilon^6) \right ] \label{eQ} \\
& & \nonumber \\
R_{_Q}^{-1} & \sim & (M^2-\omega^2)^{1/2} \approx \epsilon \, M \, \left ( 
1 +\frac{1}{2} \epsilon^2 + \frac{7}{8} \epsilon^4 + O(\epsilon^6) 
\right )
\label{size}
\eea
where $\epsilon= (Q A^2/3 S_\psi M^2) < \frac{1}{2}$ by virtue of 
the constraint (\ref{Qconstr}).  

Generalization of this discussion to the case of many scalar fields
is straightforward and involves finding the bounce in the potential
(\ref{Uhat}).  For a complicated scalar potential, as that of the MSSM, 
this can be done numerically, for example, using the Improved Action 
method~\cite{ak_n}.

\section{$B$ and $L$ balls in the MSSM}  
 
Every supersymmetric generalization of the MSSM must have Yukawa couplings 
of the Higgs fields $H_1 $ and $H_2$ to quarks and leptons which arise
from the superpotential of the form 

\beq
W=y H_2 \F \f + \tilde{\mu} H_1 H_2 +...
\label{sptn}
\eeq
Here $\F$ stands for either a left-handed quark ($\Q$), or a 
lepton ($\Lp$) superfield, and $\f$ denotes $\q$ or $\lp$, respectively.  
The corresponding scalar potential must, therefore, have cubic terms of the
form  $y \tilde{\mu} H_2 \F \f$.  In addition, there are soft supersymmetry
breaking terms of the form $y A H_1 \F \f$.  This is a generic feature of
all SSM.  

For squarks and sleptons, there are several abelian symmetries\footnote{
The case of non-abelian Q-balls associated with  
squarks and sleptons will be discussed elsewhere.} that are suitable for
building Q-balls.  These are $U(1)_{B}$, $U(1)_{{L_i}}$ and $U(1)_{E}$,  
associated with the conservation of baryon number, three types of lepton
numbers, and the electric charge.  Although we discussed only the case of a
global $U(1)$ symmetry, Q-balls can be constructed for a local $U(1)$ as
well \cite{l}.  In the case of a local symmetry, Q-balls are stable as long
as their charge is less than some maximal value~\cite{l}.

In the MSSM, Q-balls are allowed, therefore, to have a baryon number, a 
lepton number, and an electric charge.  As a toy model, one can consider a
potential for the Higgs field, $H$, and the sleptons, $\Lp$ and $\lp$, with
a scalar potential  

\beq
U=m_{_H}^2 |H|^2+m_{_L}^2 |\Lp|^2+ m_{l}^2|\lp|^2 - y A (H \Lp^* \lp
+c.c.) + y^2 (|H^2 \Lp^2| + |H^2 \lp^2| + |\Lp^2 \lp^2|) + V_{_D}, 
\label{toy}
\eeq
where $V_{_D}= (g_1^2/8) [|H|^2- |\Lp|^2]^2 +(g_2^2/8) [|H|^2+|\Lp|^2-2
|\lp|^2]^2$ 
is the contribution of the gauge the $D$-terms.   For simplicity, we 
neglected the Higgs VEV.   Nevertheless, this
toy model is instructive because it allows for some non-topological
solitons with the same quantum numbers as those in the MSSM.  The potential
is invariant under the global $U(1)_{_L}$ symmetry 
($\Lp \rightarrow \exp \{i \theta \}\Lp$ and $\lp \rightarrow \exp \{i
\theta \} \lp$) associated with the lepton 
number conservation.  Both $\Lp$ and $\lp$ have a unit charge with respect
to this $U(1)$, while the Higgs field is $U(1)_{_L}$ invariant. 

It is convenient to write

\beq
\left \{ \begin{array}{lll}
H &=& F \ sin\xi \\
\Lp &=& F \ \cos\xi \ \sin \theta \\
\lp &=& F \ \cos\xi \ \cos \theta 
\end{array} 
\right.
\label{fields}
\eeq
The condition (\ref{condmin}) is satisfied, and a Q-ball with mass 
$M_{_Q}=\mu Q$ exists,  if $\mu^2$ in equation (\ref{ElargeQ}) 
is minimized at some value of $F\neq 0$.  

\beq
\mu^2 = \frac{2U}{|\Lp|^2 + |\lp|^2} = 
\frac{1}{\cos^2 \xi} [\gamma_2(m_i^2,\xi.\theta) -
y A \gamma_3(\xi,\theta) \, F \ + \  \gamma_4 (\xi,\theta) \, F^2], 
\label{cond_gammas} 
\eeq
where $\gamma_2$ and $\gamma_4$  are non-negative functions of masses and
mixing angles, $\gamma_3= \cos^2\xi \,\sin \xi \, \sin(2\theta) $.  
The minimum of $\mu^2$ in (\ref{cond_gammas}) is achieved 
at $F \neq 0 \ $ if $\ y A \neq 0$.  The origin is not a local minimum.  
Therefore, in our toy model, $L$ balls exist no matter how small the 
tri-linear couplings might be, as long as they are non-zero.  The same is
true of the baryonic balls built of squarks.  

Of course, in the full MSSM there can be other fields that carry the same
charge.  Therefore, the local minimum of energy corresponding 
to a particular set of fields may not be the global minimum.  For example, 
an electrically  neutral selectron  $L$ ball,
$\{H,\tilde{e}_{_L},\tilde{e}_{_R} \}$, will be in competition with a 
sneutrino ball, $\{H,\tilde{\nu}_{_L},\tilde{\nu}_{_R} \}$.  However, since
the origin is not a local minimum of (\ref{cond_gammas}) for $y A \neq 0$, 
there is always a stable Q-ball with a given lepton (baryon)
number\footnote{ 
This would not necessarily  be the case if one of the sleptons or squarks
had its tri-linear coupling equal to zero (and was sufficiently light).  
However, as far as we know, this cannot happen in a realistic model, where 
the cubic couplings are allowed by the gauge symmetry, and are also
required in order to break the continious R symmetry explicitly.}. 

Having convinced ourselves that non-topological solitons exist 
in the MSSM, we will now discuss some of the phenomenological consequences.  
Large Q-balls are extended objects and cannot be produced in a collider.  
As follows from equation (\ref{size}), 
Q-beads, with charge of order a few, are also extended objects, 
whose size is large in comparison to their De Broglie wavelength.
The probability of producing them in a collider experiment is, probably,  
exponentially suppressed by their size and is likely to be negligible. 
This question, however, is by no means obvious and 
deserves a more careful analysis because, if the Q-beads can be 
created in a collider, their signatures could be spectacular.  For example,
a soliton with both $B\neq 0$ and $L\neq 0$ would interact as a
leptoquark.

In the early Universe, the non-topological solitons can be created in the
course of a phase transition \cite{fggk1,gkm} via the Kibble mechanism
(``solitogenesis''), or they can be produced in a fusion process
reminiscent of nucleosynthesis \cite{gk,gkm} (``solitosynthesis'').  Their
subsequent evolution can lead to interesting cosmological phenomena 
\cite{ellis}.  

Since the baryon and lepton
asymmetries are small (if not zero), it is the statistical fluctuations
of charge that play a major role in the formation of the baryonic and
leptonic balls.  The rate of such fluctuations was estimated in
Ref.~\cite{gkm} for a particular model. A typical soliton number to entropy
ratio was found to be $Y_{_Q}\equiv n_{_Q}/s  \sim c \, Q^{-3/2} \exp
(-Q)$, where $c$ is a  dimensionless number ($c\sim 10^{-3}$ for the model
discussed in Ref.~\cite{gkm}).  Although this estimate is expected to break
down for small $Q$, it is clear on general grounds that the  
small-charge solitons can be produced in greater numbers than the large
Q-balls.  In a separate work, we will discuss the details of the $B$ and
$L$-ball production at high temperature~\cite{wip}.  In any case, small and 
moderately large solitons can be produced in great numbers at high
temperatures in the early Universe.

Stability of very small solitons, for example those with a unit
charge\footnote{ 
Semiclassical results can be modified noticeably by quantum corrections 
if $Q=1$ \cite{ak_another}.  For instance, the soliton mass can receive 
order 1 corrections in this limit.  On the other hand, since the size of a 
$Q=1$ soliton is still large in comparison to its De Broglie wavelength 
(equation (\ref{size})), the semiclassical treatment of $Q=1$ beads 
may still be appropriate.  Since we know of no alternative 
to the semiclassical description of solitons, we will proceed keeping in
mind this caveat.},  can be guaranteed merely by some combination of the
conservation laws, regardless of the soliton mass.  For example, an
electrically neutral, $SU(2)$ singlet, $L=1$
bead with zero spin cannot decay because of the lepton number and the
angular momentum conservation.  There is simply no state in the MSSM
spectrum, except for the soliton sector, that would have these quantum
numbers.  Although caution is urged in applying the 
semiclassical treatment to Q-beads of a unit charge, there is no obvious 
reason to exclude these objects as candidates for dark matter. 

Large minimal-energy 
$B$ and $L$-balls built of squarks and sleptons can be stable against
decay into their constituent scalar fields, but they can still evaporate
into the fermions that carry $B$ and $L$, quarks and leptons \cite{ccgm}.  
According to Ref. \cite{ccgm}, the evaporation proceeds from the
surface of the Q-ball and the rate is proportional to the surface area,
rather than the volume of the Q-ball.  
This is due to the exclusion principle for fermions.  Inside the Q-ball, the
Dirac sea of quarks and leptons fills up until the Fermi pressure prevents
further production of these particles via the decay of the squarks and
sleptons.  The fermionic decay products can still leak through the surface
of the Q-ball, and the evaporation proceeds slowly, 
at the rate proportional to the surface area.  The evaporation rate would
be proportional to the volume of the Q-ball if it were to decay into scalar
particles.  However, we saw that this is forbidden by the
energy conservation for the Q-balls of minimal energy. Gauge fields carry
no $(B-L)$ charge and cannot facilitate the evaporation. 

In the MSSM, the processes that can lead to $B$ and $L$ balls evaporation 
into quarks and leptons are mediated by gauginos (and gluinos) and, if the
gaugino mass is larger than $\mu$, they can be further suppressed. 
The lifetimes of baryonic and leptonic balls built of
sparticles are model-dependent and will be analysed elsewhere~\cite{wip} 
for a variety of the MSSM parameters. 

Those $B$ and $L$ solitons that decay at temperatures $T$ above 1~GeV, 
probably, have no observable consequences.  However, a remarkable 
transformation can take place for a Q-ball that survived to a temperature of
order $\Lambda_{QCD}$.  We recall that the interior of a large evaporating
Q-ball is populated with a high density of quarks that fill the Dirac sea
up to the energies of order $\mu$.  If the Q-ball survives to temperatures
below $\Lambda_{QCD}$, then the population of quarks fostered inside the 
sparticle ball can remain bound, now by the $QCD$ forces, even after the 
sparticle structure, which kept them together originally, disappears.  
At $T \gg 1$~GeV, such a conglomerate of nuclear matter would thermalize
without a trace.  However, at lower temperatures, heavy nuclei can form as 
vestiges of sparticle Q-balls. 

Since the statistical fluctuation mechanism \cite{gkm}  is probably the
most likely source of solitons at the electroweak scale temperatures, a
comparable numbers of baryon (lepton) and anti-baryon (anti-lepton) balls 
will be produced.  Those Q-balls that have a lifetime of order $10^{-6}$~s
or more, will give birth to heavy nuclei ($A \sim Q$) of matter and
anti-matter, with some excess for $B>0$.  The excess of $B>0$ nuclei can
survive the subsequent annihilation.

This allows for a highly non-standard synthesis of heavy nuclei in the 
early Universe, such that they are already present at the time $t\sim 1$~s,
when the standard nucleosynthesis is supposed to commence.  Fission of
heavy nuclei can also be the source of additional lighter elements, in
particular, $^4He$, which are copiously produced in nuclear decays. 
Details of this and other cosmological implications of the MSSM 
solitons will be analyzed elsewhere~\cite{wip}. 

Q-balls with lifetime longer than 1~second are probably disallowed, at
least if they can be produced in substantial quantities. Their decay
products can cause an unacceptable increase in entropy, or disturb
the spectrum of the microwave background radiation.  

In summary, non-topological solitons with non-zero baryon and lepton
number, as well as the electric charge, are  generically present in the
spectrum of the MSSM and other models with low-energy supersymmetry. 
Production of these objects in the early Universe can have a number of
important cosmological ramifications.

The author would like to thank S.~Dimopoulos and M.~Shaposhnikov 
for many interesting discussions and L.~\'Alvarez-Gaum\'e, G.~Dvali,
S.~Lola and G.~Veneziano for helpful conversations.

\end{document}